\documentclass[conference]{IEEEtran}
\IEEEoverridecommandlockouts
\usepackage{cite}
\usepackage{amsmath,amssymb,amsfonts}
\usepackage{algorithmic}
\usepackage{graphicx}
\usepackage{textcomp}
\usepackage[T1]{fontenc}
\usepackage[hyphens]{url}
\usepackage{array,booktabs,tabularx}
\usepackage[table]{xcolor}
\usepackage{tikz}
\usetikzlibrary{arrows.meta,patterns.meta}
\usepackage{hyperref}\hypersetup{hidelinks}
\usetikzlibrary{positioning,arrows.meta}
\definecolor{ink}{HTML}{24445A}
\definecolor{native}{HTML}{176C74}
\definecolor{plugin}{HTML}{8A5C2D}
\renewcommand{\arraystretch}{1.12}
\definecolor{groupbg}{gray}{0.92}
\newcommand{\tabgroup}[2]{\rowcolor{groupbg}[0pt][0pt]%
  \multicolumn{#1}{@{}c@{}}{\rule[-0.6ex]{0pt}{2.5ex}\emph{#2}}\\}
\def\BibTeX{{\rm B\kern-.05em{\sc i\kern-.025em b}\kern-.08em
    T\kern-.1667em\lower.7ex\hbox{E}\kern-.125emX}}
\begin{document}

\title{When Routine Operations Break Architectural Pattern Relationships in LLM Agent Harnesses}
\author{\IEEEauthorblockN{Boming Xia\IEEEauthorrefmark{1}\IEEEauthorrefmark{2}\IEEEauthorrefmark{3},
Qinghua Lu\IEEEauthorrefmark{3}\IEEEauthorrefmark{4}\IEEEauthorrefmark{1},
Dino Sejdinovic\IEEEauthorrefmark{1}\IEEEauthorrefmark{2} and
Liming Zhu\IEEEauthorrefmark{3}\IEEEauthorrefmark{4}}
\IEEEauthorblockA{\IEEEauthorrefmark{1}Responsible AI Research Centre \IEEEauthorrefmark{2}Adelaide University \IEEEauthorrefmark{3}CSIRO \IEEEauthorrefmark{4}UNSW}}
\hyphenation{Deep-Seek}
\maketitle
\raggedbottom
\providecommand{\dg}{\textsuperscript{*}}
\begin{abstract}
LLM agent harnesses distribute tool mediation, state management and safety guardrails across native services, independently written plugins and configuration. Routine operations can change the runtime order or retained state that architectural pattern relationships require, without any code change. We study five relationships associated with the Guardrail Enforcement Point, Autonomy Controller and Blackboard patterns across five released harnesses, analysing a main sample of fourteen independently written plugins. Using PatternOps, our assessment procedure, we trace requirements from pattern descriptions, provider documentation and deployment profiles to the providers responsible for them and the records used to assess them. We then compare alternative allocations under supported operations. After a policy plugin was reapplied or an edit allow rule was saved, a prohibited edit executed without policy evaluation even though the participating plugins were active. Which alternatives qualified depended on the requirement. Allocations that kept the required order, or kept or rebuilt the needed state, held on their tested paths and operation sequences, while others prevented prohibited effects but missed edits rejected elsewhere. In these harnesses, a pattern relationship is a property of the running system and its history, so architects should evaluate it against the decision paths, state lifetimes and operation sequences their deployments use.
\end{abstract}

\section{Introduction}\label{sec:intro}
An LLM agent harness can execute a prohibited edit while its policy and approval plugins remain active. Harnesses coordinate model and tool calls, manage execution state and mediate actions through mechanisms such as policy checks and operator approval~\cite{msagentharness2026,googleagentharness2026,barbaste2026,liu2026claudecode,asd2026harnesses}. Architectural patterns describe reusable arrangements of responsibilities and relationships for concerns such as guardrail enforcement, operator control and shared state~\cite{xia2026agentops,ufoblackboard2026,nii1986blackboard}. Native services, independently written plugins and configuration realise these responsibilities jointly, so their runtime interactions decide whether the required relationships survive routine operations, such as reapplying a plugin, saving consent or compacting the history.\looseness=-1

\begin{figure*}[t]\centering
% Figure 1: grayscale / print-friendly TikZ picture
% Requires:
%   \usepackage{tikz}
%   \usetikzlibrary{arrows.meta,patterns.meta}

% Figure 1: print-friendly grayscale rendering; no external styles required.
% Preamble: \usepackage{tikz}
%           \usetikzlibrary{arrows.meta,patterns.meta}
% Note: do not insert blank lines inside the TikZ option list below.
\definecolor{figInk}{gray}{0.20}
\definecolor{figPanel}{gray}{0.92}
\definecolor{figLite}{gray}{0.97}
\definecolor{figDark}{gray}{0.76}
\begin{tikzpicture}[
  x=.98cm,y=1cm,>=Stealth,font=\footnotesize,
  every node/.append style={text=figInk},
  box/.style={rounded corners=2pt,align=center,text width=23mm,
    minimum height=7.8mm,inner xsep=2.5pt,inner ysep=1.5pt},
  % Role styles: the same definitions are reused by process and legend nodes.
  nativeRole/.style={draw=figInk!65,fill=figLite,line width=.48pt},
  pluginRole/.style={draw=figInk!65,fill=white,
    pattern={Lines[angle=45,distance=5pt,line width=.18pt]},
    pattern color=black!28,line width=.48pt},
  policyRole/.style={draw=figInk!65,fill=white,
    pattern={Dots[distance=4.0pt,radius=.34pt]},
    pattern color=black!48,line width=.48pt},
  humanRole/.style={draw=figInk!65,fill=white,line width=.48pt},
  heldRole/.style={draw=figInk,fill=figDark,line width=.85pt},
  lossRole/.style={draw=figInk,fill=white,
    pattern={Hatch[angle=45,distance=6.5pt,line width=.16pt]},
    pattern color=black!22,line width=.85pt},
  absentRole/.style={draw=figInk!45,dashed,fill=white,
    text=figInk!80,line width=.48pt},
  nativebox/.style={box,nativeRole},
  pluginbox/.style={box,pluginRole},
  policybox/.style={box,policyRole},
  humanbox/.style={box,humanRole},
  heldbox/.style={box,heldRole},
  lossbox/.style={box,lossRole},
  absentbox/.style={box,absentRole,font=\scriptsize},
  link/.style={-{Stealth[length=1.65mm]},draw=figInk,semithick},
  lab/.style={anchor=west,align=left,text width=32.5mm,inner sep=0pt},
  mini/.style={rounded corners=2pt,minimum height=4.1mm,
    text width=19mm,inner xsep=2.5pt,align=center,font=\scriptsize}
]

% ------------------------------------------------------------------
% Framed panels
% Slightly reduced panel width and shifted last-column nodes right
% to reduce excess blank space on the right.
% ------------------------------------------------------------------

\path[draw=figInk!42,fill=white,rounded corners=3pt,line width=.45pt]
  (0,.12) rectangle (15.30,-4.40);
\begin{scope}
  \clip[rounded corners=3pt] (0,.12) rectangle (15.30,-4.40);
  \fill[figPanel] (0,.12) rectangle (15.30,-.46);
\end{scope}

\path[draw=figInk!42,fill=white,rounded corners=3pt,line width=.45pt]
  (0,-4.53) rectangle (15.30,-7.64);
\begin{scope}
  \clip[rounded corners=3pt] (0,-4.53) rectangle (15.30,-7.64);
  \fill[figPanel] (0,-4.53) rectangle (15.30,-5.04);
\end{scope}

\foreach \top/\bottom in {-.67/-1.53,-2.20/-3.09,-3.31/-4.23,-5.20/-6.12,-6.54/-7.48} {
  \path[fill=figInk!3,rounded corners=2pt] (.20,\top) rectangle (3.67,\bottom);
}

% ------------------------------------------------------------------
% (a) DeepSeek Harness
% ------------------------------------------------------------------

\node[anchor=west,font=\footnotesize\bfseries,text=figInk] at (.32,-.19)
  {(a)\quad Callback order on DeepSeek Harness rc.2};

\node[lab] at (.39,-1.09) {\textbf{Arranged policy first}\\(observe mode)};
\node[pluginbox] (a11) at (4.85,-1.09) {Governance\\submits call};
\node[policybox] (a12) at (7.62,-1.09) {Policy records\\decision};
\node[pluginbox] (a13) at (10.39,-1.09) {Tool Approval\\asks};
\node[humanbox]  (a14) at (13.45,-1.09) {Operator\\decides};

\draw[link] (a11.east)--(a12.west);
\draw[link] (a12.east)--(a13.west);
\draw[link] (a13.east)--(a14.west);

\path[fill=figPanel!70,rounded corners=1.5pt]
  (.22,-1.63) rectangle (15.07,-2.12);
\node[anchor=west,font=\footnotesize\itshape,text=figInk] at (.47,-1.88)
  {\textbf{Mode switch (observe $\to$ enforce):} Governance re-registers behind Tool Approval};

\node[lab] at (.39,-2.65) {\textbf{After the switch}\\Prohibited edit};
\node[pluginbox] (a21) at (4.85,-2.65) {Tool Approval\\ends chain};
\node[humanbox]  (a22) at (7.62,-2.65) {Operator\\approves};
\node[lossbox]   (a23) at (10.39,-2.65) {Prohibited edit\\executes};
\node[absentbox] (a24) at (13.58,-2.65) {Not reached:\\Governance / policy};

\draw[link] (a21.east)--(a22.west);
\draw[link] (a22.east)--(a23.west);

\draw[figInk!22,thin] (.35,-3.20)--(14.96,-3.20);

\node[lab] at (.39,-3.76) {\textbf{Alternative}\\Native consent};
\node[pluginbox] (a31) at (4.85,-3.76) {Governance\\submits call};
\node[policybox] (a32) at (7.62,-3.76) {Policy\\blocks};
\node[heldbox]   (a33) at (10.39,-3.76) {No edit\\applied};
\node[absentbox] (a34) at (13.58,-3.76) {Not asked:\\native question};

\draw[link] (a31.east)--(a32.west);
\draw[link] (a32.east)--(a33.west);

% ------------------------------------------------------------------
% (b) OpenCode
% ------------------------------------------------------------------

\node[anchor=west,font=\footnotesize\bfseries,text=figInk] at (.32,-4.80)
  {(b)\quad Saved consent and Permission Reviewer on OpenCode 2.2.5};

\node[lab] at (.39,-5.65) {\textbf{Control arm}\\No saved allow rule};
\node[nativebox] (b11) at (4.85,-5.65) {Permission stage:\\would ask};
\node[pluginbox] (b12) at (7.62,-5.65) {Reviewer\\evaluates};
\node[policybox] (b13) at (10.39,-5.65) {Policy\\denies};
\node[heldbox]   (b14) at (13.45,-5.65) {No edit\\applied};

\draw[link] (b11.east)--(b12.west);
\draw[link] (b12.east)--(b13.west);
\draw[link] (b13.east)--(b14.west);

\path[fill=figPanel!70,rounded corners=1.5pt]
  (.22,-6.19) rectangle (15.07,-6.50);
\node[anchor=west,font=\footnotesize\itshape,text=figInk] at (.47,-6.34)
  {\textbf{Matched arms:} An earlier ``always'' reply saved a rule allowing edits to any file in the saved-rule arm};

\node[lab] at (.39,-7.01) {\textbf{Saved-rule arm}\\Prohibited edit};
\node[nativebox] (b21) at (4.85,-7.01) {Permission stage:\\already allowed};
\node[pluginbox] (b22) at (7.62,-7.01) {Reviewer abstains\\(not an ask)};
\node[lossbox]   (b23) at (10.39,-7.01) {Prohibited edit\\executes};
\node[absentbox] (b24) at (13.58,-7.01) {No policy\\request};

\draw[link] (b21.east)--(b22.west);
\draw[link] (b22.east)--(b23.west);

% ------------------------------------------------------------------
% Right-hand key
% ------------------------------------------------------------------

\path[draw=figInk!42,fill=white,rounded corners=3pt,line width=.45pt]
  (15.56,.12) rectangle (18.05,-7.64);
\begin{scope}
  \clip[rounded corners=3pt] (15.56,.12) rectangle (18.05,-7.64);
  \fill[figPanel] (15.56,.12) rectangle (18.05,-.46);
\end{scope}

\node[anchor=center,font=\footnotesize\bfseries,text=figInk] at (16.80,-.19) {Key};

\node[mini,nativeRole] at (16.80,-1.00) {Native stage};
\node[mini,pluginRole] at (16.80,-2.00) {Plugin};
\node[mini,policyRole] at (16.80,-3.00) {Study policy};
\node[mini,humanRole]  at (16.80,-4.00) {Operator};

\draw[figInk!25,thin] (15.85,-4.52)--(17.75,-4.52);

\node[mini,heldRole]   at (16.80,-5.05) {Edit prevented};
\node[mini,lossRole]   at (16.80,-6.05) {Edit executed};

\draw[figInk!25,thin] (15.85,-6.53)--(17.75,-6.53);

\node[mini,absentRole] at (16.80,-7.05) {Off-path};

\end{tikzpicture}
\caption{Policy participation under supported operations. (a) A mode switch reorders callbacks in DeepSeek Harness. \emph{Native consent}, the host's own approval question after the plugin chain, keeps policy reachable in enforce mode under the tested read-only profile. (b) A saved allow rule makes OpenCode's reviewer ineligible. Rows of (b) are matched arms in separate processes.}\label{fig:providers}
\end{figure*}
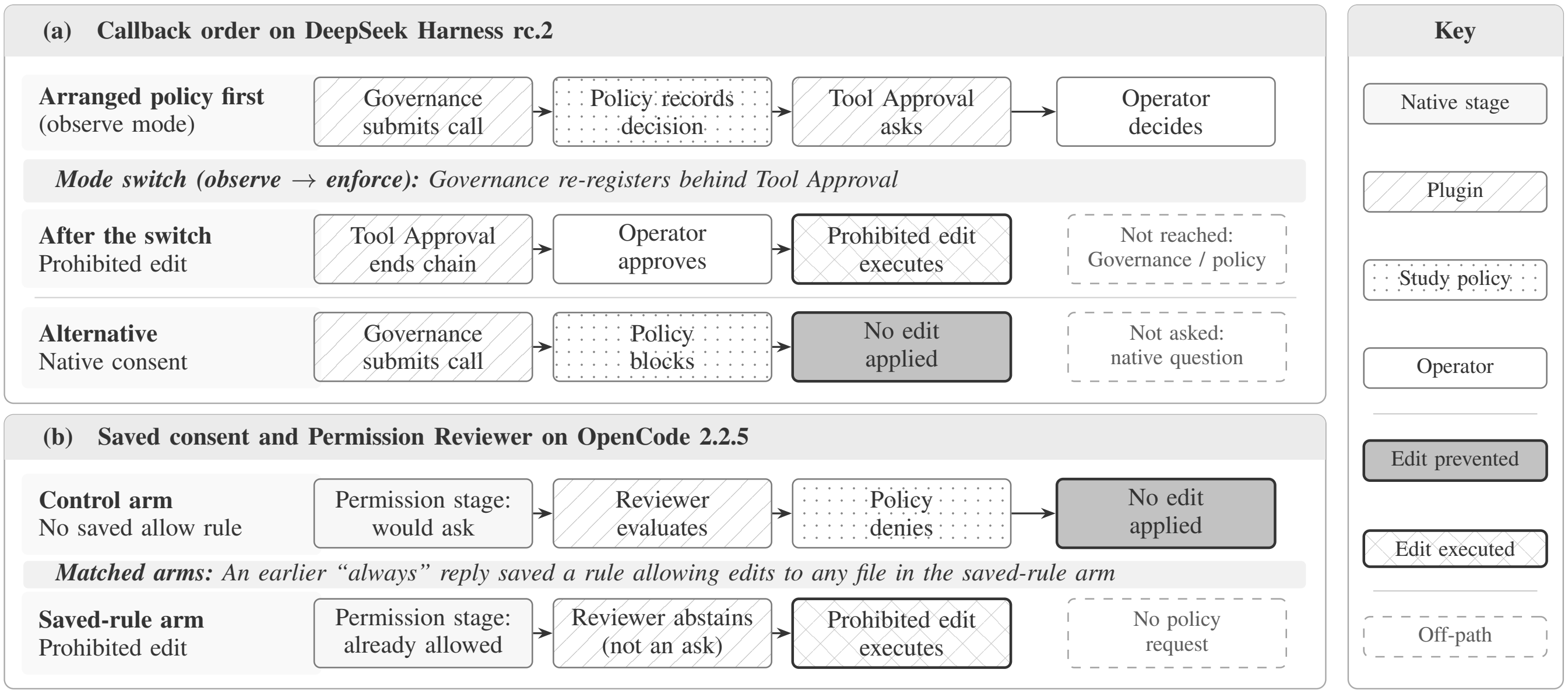

Consider \emph{policy participation}, an assessed Guardrail Enforcement Point (GEP) requirement: the designated policy must decide before an action \emph{covered} by the deployment executes, even when an operator approves it. In one harness, a policy plugin submits tool calls to a policy service and an approval plugin asks the operator~\cite{deepseekharness2026,dshgovernance2026,dshtoolapproval2026}. Their callbacks form a chain of responsibility~\cite{gamma1994}: by default, they run in registration order until one answers without passing the call on~\cite{deepseekharness2026}. We arranged policy-first order and confirmed that an initial edit in observe mode, used for calibration, reached policy. The documented switch to enforce mode re-registered the policy callback behind the approval plugin. The operator then approved a prohibited edit, and it executed without a policy decision. Disabling and re-enabling the approval plugin re-registered it last and restored evaluation. The same enabled plugins and final settings thus gave different policy coverage after different operations (Fig.~\ref{fig:providers}a).\looseness=-1

Runtime order is one way to carry a relationship. Retained state is another: after an allow rule was saved, a policy reviewer that checks only requests still needing operator approval abstained, and a prohibited edit executed~\cite{opencode2026,permreviewer2026} (Fig.~\ref{fig:providers}b). Reapplying a plugin can also discard source provenance held in its instance. An \emph{allocation} assigns responsibilities to harness parts, and a \emph{carrier} is the runtime order or retained state on which a required relationship depends. Preservation through routine operations therefore depends on the allocation as a whole, not on any plugin in isolation.\looseness=-1

These dependencies reflect established concerns in complete mediation, feature interaction, aspect precedence, safe dynamic update and architectural mismatch~\cite{saltzer1975protection,calder2003,douence2004,kramer1990,hayden2012kitsune,garlan1995mismatch}. Recent studies examine runtime pattern conformance, interacting changes to agent harnesses, action-bound approval and contracts or shared decision cores for composed agent controls~\cite{krieger2020,zhang2026csf,kumar2026loopjacking,zheng2026continuity,zhang2026agate}. For an architect choosing where to enforce policy or retain decision state, the remaining question is which alternative allocations among a harness's native services and independently written plugins preserve a specified relationship through the operations a deployment supports, and what obligations each allocation introduces. Answering it requires knowing what carries each relationship and how operations and routes change it.\looseness=-1

We investigate five assessed relationships across five released harnesses as a multiple-case study with controlled comparisons~\cite{runeson2009}, analysing a main sample of fourteen independently written plugins from source. GEP is the principal case, covering policy participation and \emph{decision information}, such as provenance needed for a policy decision. Supporting comparisons examine whether an Autonomy Controller (AC) response applies to its pending action, whether resumption repeats completed work under a task-derived requirement and whether a shut-down Blackboard member remains excluded under a specified deployment profile. Using PatternOps, our assessment procedure (Section~\ref{sec:design}), we distinguish pattern duties, provider promises and deployment requirements. We trace each requirement to its allocation and carrier and judge the assessed actions from records joined across providers\footnote{The companion and anonymised evidence are available at \url{https://anonymous.4open.science/r/pattern-relationships-supplement-C35A/}.}. We ask:

\begin{itemize}\interlinepenalty=10000
\item \textbf{RQ1:} How are the assessed relationships realised across native capabilities, plugins and configuration, and what carries them?
\item \textbf{RQ2:} How do supported operations and execution routes preserve or break these relationships?
\item \textbf{RQ3:} Which alternative allocations preserve them under the tested conditions, and with what trade-offs?
\end{itemize}

The study contributes:
\begin{enumerate}
\item An evidence-grounded account of how selected architectural relationships are allocated and carried across released agent harnesses (RQ1).
\item Controlled evidence of operation-dependent losses and restorations, including different policy coverage with the same enabled plugins and final settings, and a state repair that fails after a subsequent operation (RQ2).
\item Requirement-specific comparisons of alternative allocations, distinguishing prevention of prohibited effects from evaluation of every admitted edit, together with their tested conditions and operating obligations (RQ3).
\end{enumerate}

\section{Requirements and carriers}\label{sec:background}
Judging whether a relationship holds needs two distinctions: who sets the requirement it must meet, and what runtime order or state carries it. The harness core is the \emph{host}. An \emph{assembly} combines a particular host release with its plugins, configuration and services. The parts of an assembly that supply a pattern's responsibilities are that pattern's \emph{providers}, and a \emph{pattern relationship} is a required relation among the responsibilities of one pattern, realised by one or several providers. Each judgment concerns one pattern relationship and the requirements stated for it (Table~\ref{tab:requirements}), not the whole pattern.\looseness=-1

\textbf{Authority.} Requirements have three authorities: pattern descriptions supply \emph{pattern duties}, provider documentation supplies \emph{provider promises} and the study's deployment profile supplies \emph{deployment requirements}. A provider can meet its promise while its allocation fails another requirement. Table~\ref{tab:requirements} records the authorities and the fixed \emph{study readings}. Policy participation has three nested requirements. \emph{Effect prevention}, the weakest deployment requirement, needs only that no prohibited effect occurs. The GEP pattern duty adds that the designated policy decides before a covered action executes and that its prohibitions are enforced. Calibrating policy on all admitted proposals requires it to evaluate even edits that another provider rejects. We call this stricter deployment requirement the \emph{every-edit requirement}. An edit that another provider rejects before policy sees it thus violates only the every-edit requirement. An edit that executes unevaluated also violates the GEP duty, and a prohibited one also violates effect prevention. Policy can also escalate a call to the operator for consent. We report escalation handling separately. Every policy-participation comparison also requires that an operator's rejection stops the call it answers.\looseness=-1

\begin{table*}[t]
\centering\small
\caption{Assessed relationships and requirement authorities.}
\label{tab:requirements}
\setlength{\tabcolsep}{3pt}
\renewcommand{\arraystretch}{0.97}
\begin{tabularx}{\textwidth}{@{}>{\raggedright\arraybackslash}p{.21\textwidth}>{\raggedright\arraybackslash\everypar{\hangindent=1em\hangafter=1}}X@{}}
\toprule
\textbf{Relationship} & \textbf{Authority of assessed requirement} \\
\midrule
\textbf{GEP: policy participation} &
\textbf{Pattern duty:} The inline enforcement point ``must return before execution proceeds'' and enforces prohibitions returned by policy (\S3.2.1).\par
\textbf{Provider promise:} Governance ``intercepts every tool call''.\par
\textbf{Deployment requirement:} A study policy service decides; approval does not override a prohibition; every admitted edit call is evaluated, including rejected ones (the every-edit requirement). \\
\cmidrule{1-2}
\textbf{GEP: decision information} &
\textbf{Pattern duty:} The decision receives the candidate action and its boundary context.\par
\textbf{Provider promise:} Airlock: untrusted content cannot direct an egress-capable tool, and compaction cannot launder (remove) a label.\par
\textbf{Deployment requirement:} An untrusted source restricts edits, or under Airlock's built-in rules egress-capable tools, while it is in the model's input or a summary's sources. RiskProof: an observe-mode read restricts the first enforce decision. \\
\cmidrule{1-2}
\textbf{AC: response applied} &
\textbf{Pattern duty:} The operator's response is applied before progression (\S3.2.2). \emph{Study reading:} the response to the pending action. \\
\cmidrule{1-2}
\textbf{AC: completed work} &
\textbf{Deployment requirement (task):} Resumption repeats no completed work. \\
\cmidrule{1-2}
\textbf{Blackboard: membership} &
\textbf{Pattern duty:} Knowledge is ``accessible to all agents''. \emph{Study reading:} participating members.\par
\textbf{Provider promise:} Ensemble: its lookup returns only members that have not shut down.\par
\textbf{Deployment requirement:} A shut-down member stays excluded until it re-enrols. \\
\bottomrule
\end{tabularx}
\par\smallskip
\parbox{\textwidth}{\footnotesize
Sources: the AgentOps catalogue for GEP and AC (\S\ numbers)~\cite{xia2026agentops}, Microsoft's agent memory design for Blackboard~\cite{ufoblackboard2026} and Nii's model as background~\cite{nii1986blackboard}. Airlock and RiskProof track source provenance in DeepSeek Harness, and Ensemble is an OpenCode board plugin.}
\end{table*}

\textbf{Carriers and operations.} Carriers come in two kinds, and one relationship can depend on both. A connector's order or position, such as the callback chain, decides whether policy sees an edit before an approval question ends the call. Retained state, such as saved consent, a plugin's private labels or a member cache, decides which requests a provider acts on and what a decision still knows. Carriers thus lie in a connector's semantics~\cite{mehta2000connectors} or in retained state. The studied operations are ones the hosts support for normal use, and each can change a carrier without changing any provider's code. On DeepSeek Harness, \emph{reapplication} disposes of an instance and creates another after its configuration changes, as the switch from observe to enforce mode does for the policy plugin. A \emph{reload}, here on OpenCode and Pi, replaces instances without a configuration change. Both thus replace a component instance in the sense of dynamic reconfiguration~\cite{kramer1990}. \emph{Compaction} replaces an earlier interval of the conversation history with a summary and keeps a later tail verbatim.\looseness=-1

\section{Study design}\label{sec:design}
The cases were chosen purposively to expose different ways in which harnesses compose decision providers and retain their state. Source analysis and observation on the released hosts locate each relationship's providers and carrier (RQ1). Comparisons then vary operations and execution routes (RQ2) and allocations (RQ3) against the requirements of Section~\ref{sec:background}.

\textbf{PatternOps.} PatternOps is the specified procedure this study uses to turn a pattern relationship into requirements that an assembly's own records can decide (companion Section~III). The inputs are a fixed pattern description, an identified assembly and an assessment question. Assessors record what they knew in advance, derive the pattern duties and identify competing readings of ambiguous phrases. Source authority settles the reading. They record provider promises, fix deployment requirements and bind each requirement to its providers and carrier in a Pattern Realisation Contract. Each pattern and host gets its own derivation and bindings.

\textbf{Judgments.} Assessors join each action's records across providers and label each requirement part satisfied, violated, unsupported (the records decide neither way), not exercised (the boundary was not reached) or not applicable. Within a fixed requirement and judgment scope, labels for applicable parts combine in this order: violated, unsupported, not exercised then satisfied. An unsupported part alone never establishes a violation. If an inspected path lacks the required decision but an uninspected provider could have supplied it, the path judgment is violated and the assembly judgment unsupported. A positive judgment must hold under every reading fixed for the assessment. Later readings are reported as sensitivities.\looseness=-1

\textbf{Subjects.} The main sample comprises fourteen independently written plugins analysed from source (companion Section~IV lists them): eleven for DeepSeek Harness and three for OpenCode. Exploratory and bounded repository searches selected the DeepSeek Harness plugins for different control and state arrangements. From code, six, including Governance and Airlock, have approval branches that end the chain without delegating (passing the call on) for the tools they cover, so each can skip a policy callback registered after it. The OpenCode subjects are the policy reviewers Permission Reviewer and ACP and the board plugin Ensemble~\cite{permreviewer2026,acp2026,ensemble2026}. ACP was selected after the saved-rule loss, because its code also reviews already-allowed requests.

Host comparisons exercise ten sample plugins at the pinned releases of Table~\ref{tab:overview}~\cite{deepseekharness2026,opencode2026,deepagents2026,claudecodehooks2026,pi2026}. \emph{Transfer cases} substitute Intutic for the policy plugin Governance~\cite{intutic2026} and Edit Approval or AgentGuard for the approval plugin Tool Approval~\cite{dsheditapproval2026,agentguard2026}. Edit Approval runs on the host's earlier 0.1.0-rc.8 release, and AgentGuard's approval branch delegates. The provenance plugins Airlock and RiskProof supply decision information~\cite{dshairlock2026,dshriskproof2026}. On Deep Agents the AC subjects are LangChain's batched and per-call review, including Deep Agents' own approval option, which builds batched review. The vendors' Hookify (Claude Code) and plan-mode (Pi) plugins supply comparisons outside the main sample~\cite{hookify2026,pi2026}. A bounded search for repositories created after the main study selected one further OpenCode plugin, opencode-guard, as a boundary case~\cite{opencodeguard2026}.

Baseline subjects keep their code unchanged. Tables mark each provider native (N), external (E, an existing plugin loaded separately from the host) or study-built (S, code the study wrote or modified). Below, the policy plugin means Governance, the approval plugin Tool Approval, the provenance plugin Airlock and the board plugin Ensemble.\looseness=-1

\begin{table}[!t]\centering\small
\caption{Hosts, carriers and tested operations. GEP is the principal case.}\label{tab:overview}
\setlength{\tabcolsep}{4pt}\renewcommand{\arraystretch}{0.95}
\newcolumntype{H}[1]{>{\raggedright\arraybackslash\hangindent=0.8em\hangafter=1}#1}
\begin{tabularx}{\linewidth}{@{}H{p{.35\linewidth}}H{X}@{}}\toprule
\textbf{Host} & \textbf{Carrier: operations or routes}\\\midrule
\tabgroup{2}{GEP policy participation}
DeepSeek Harness rc.2, alpha.2, rc.5, rc.8 & Callback order: start, switch, configuration edit, disable and enable\\
OpenCode 2.2.5 & Permission stage under saved consent: reload, saved allow rule, revoke\\
Claude Code 2.1.317 & Hook precedence: saved allow rule\\\addlinespace[2pt]
\tabgroup{2}{GEP decision information}
DeepSeek Harness rc.2, alpha.2, rc.5 & Provenance labels in a plugin instance or the history: reapplication, compaction, compaction then reapplication\\
Pi 1.5.0 & Session log: reload, resume\\\addlinespace[2pt]
\tabgroup{2}{AC response applied}
Deep Agents 0.7.28, 0.7.13 & Route to the tool node: normal and direct routes. Pending request: graph recreation\\
OpenCode 1.19.3 & Pending request: reattachment\\\addlinespace[2pt]
\tabgroup{2}{AC completed work (task-derived)}
Deep Agents 0.7.28 & Saved task results: resumption after approval or rejection\\\addlinespace[2pt]
\tabgroup{2}{Blackboard membership}
OpenCode 1.18.30, 2.0.32 & Member cache and stored status: running lookup, restart, remove and restore\\\bottomrule
\end{tabularx}
\par\smallskip\parbox{\linewidth}{\footnotesize DeepSeek Harness rc.2, alpha.2 and rc.5 are 0.2.1 prereleases, and rc.8 is 0.1.0-rc.8. Operations are pooled across the listed releases.}
\end{table}

\textbf{Comparisons.} Matched comparisons fix a relationship's requirement and workload and, for GEP, its policy, and vary one allocation or one operation. Programmed comparisons also fix the action sequence, whereas live ones fix the task and let the model choose its actions. A changed allocation can bundle several mechanisms, so a contrast between allocations measures the allocation as a whole. A \emph{provider variant}, such as a repackaged plugin, changes how one provider is built without moving a responsibility. The execution unit is a \emph{process} (one fresh start of the host) or, for some comparisons, a \emph{workflow} (a planned run that may include a restart or graph recreation).\looseness=-1

\textbf{Execution.} Programmed model responses and operator replies drive native dispatch, approval, storage and tools. They fix the action sequence. For the five-edit DeepSeek Harness workload, a deterministic study policy service decides by tool and file name. The programmed model returns identical summaries at compaction. Most DeepSeek Harness operations use a study controller in the host process. Selected runs use \emph{public operations alone}: an outside client drives the stock command-line interface (CLI) and plugin management, with no study controller or recorder in the host.

Across seven DeepSeek Harness and OpenCode comparisons, one \emph{CLI model} and five API models replace the programmed model, so that models choose their own actions (companion Section~VIII). The CLI model's serving identity was unreported. An adapter checks the response format before forwarding final text or one call to an offered tool. Dispatch, policy, consent and effects run in the assembly. Two maintenance processes used the host's own model route without the adapter.\looseness=-1

\textbf{Units and controls.} An attempt counts once the host admits it as a tool call within the covered scope: the selected edit tools and, in the maintenance task, also shell tools. Unless stated otherwise, DeepSeek Harness policy-participation tests propose five selected edits after the operation under test, with the approval plugin limited to edits. The policy would block one edit, allow two and escalate two. The operator approves the prohibited edit if asked, approves one edit in each remaining pair and rejects the other.

Controls check policy activity, legitimate progress and relevant state before or after the operation, as each comparison requires. On DeepSeek Harness an allowed and a blocked read establish policy activity on the read path, judged separately from edits. A legitimate action, one that policy allows, must proceed when approved and stop at a \emph{matching} rejection, the operator's reply to that call's own question. A provider of decision information must restrict edits, or egress-capable tools under Airlock's built-in rules, while a fresh untrusted source is relevant. A continuing Blackboard member must receive board items.\looseness=-1

\textbf{Joined records and evidence.} An absent effect does not establish policy participation, since another provider can also prevent it. The unit of judgment is therefore one relationship at one action or state transition, read with the history it needs. The companion specifies each relationship's deciding join. Repeated programmed units reproduced their outcomes, and other releases, packaging and clients, public operations alone and live models test sensitivity to those conditions. Inputs and expectations, informed by the pinned code and earlier results, were bound before execution for the initial DeepSeek Harness mode-switch comparisons, Deep Agents' approval option, Ensemble 0.20.0 and the later-selected opencode-guard. \lq\lq From code\rq\rq\ marks claims, including explanations of observed outcomes, that rest on reading the pinned code alone. Asterisks identify results based on analyses corrected after collection on unchanged records (companion Section~IX). For RQ3 the study records added attempts and questions, changed decision behaviour and operating obligations separately.\looseness=-1

\textbf{Worked assessment.} Take the prohibited edit after the switch on DeepSeek Harness rc.2 (Fig.~\ref{fig:providers}a). The GEP duty requires the designated policy to decide before a covered action executes (Table~\ref{tab:requirements}), and the study policy prohibits this edit. Governance submits calls to that policy, Tool Approval asks the operator and the host orders their callbacks. The admitted call joins to the operator's approval and the changed file but to no policy record with the same session, tool and arguments. The callback inventory places Governance behind Tool Approval, whose answer ends the chain, and a blocked read in the same process shows that policy stayed active on the read path. Without records covering the edit path, the missing policy record would leave participation unsupported. Here the captured call, approval, effect and complete policy stream establish violations of all three nested requirements.\looseness=-1

\section{Policy participation}\label{sec:participation}
Two carriers decided policy participation: callback order on DeepSeek Harness and saved consent at OpenCode's permission stage. On both hosts an ordinary operation changed the carrier and opened a path around policy with the participating plugins active (Fig.~\ref{fig:providers}).

\textbf{Starts put approval first despite the configured row order.} DeepSeek Harness runs plugin callbacks in registration order by default, and no declared rule sets that order: the host's base profile, which lists one plugin per row, states that \lq\lq Row order carries no load semantics\rq\rq. In all 24 inventoried rc.2 and alpha.2 starts of the unchanged pair that recorded callback order, the approval plugin registered first, although the configuration lists the policy plugin first. From code, the loader registers each plugin when its own import completes, so packaging can affect the order, which the repackaging comparison below tests. In cold observe-mode starts that tested edits before any rearrangement, and in fresh starts from the final enforce settings, policy evaluated no selected edit (Table~\ref{tab:participation}).\looseness=-1

\textbf{The documented switch undid a policy-first order.} For the unchanged pair, we arranged policy-first order. Policy then evaluated every selected observe-mode edit. The switch reapplies the policy plugin. Reapplication disposes of its registration and appends a new one, so its new callback sat behind the approval plugin's, as the ordered callback inventory shows. Policy then evaluated no selected edit, and in each process the edit that policy prohibits executed once the operator approved it (Table~\ref{tab:participation}). The loss recurred across releases, including rc.5, and packaging, through public operations alone and for live models' own edits. Under the approval plugin, the five API models made 28 reads and 28 edits, and policy evaluated every read and no edit, at startup or after the switch.\looseness=-1

\begin{table}[t]
\centering\small
\caption{Selected edit attempts evaluated by policy, by allocation and operation.}
\label{tab:participation}
\setlength{\tabcolsep}{4pt}\renewcommand{\arraystretch}{0.95}
\textbf{(a) DeepSeek Harness rc.2}\par\smallskip
\begin{tabularx}{\columnwidth}{@{}>{\raggedright\arraybackslash}Xc@{}}
\toprule
\textbf{Condition / comparison} & \textbf{Evaluated / edits}\\
\midrule
\tabgroup{2}{Governance + Tool Approval}
Cold observe start & 0/1\\
Policy-first observe $\rightarrow$ switch & 1/1 $\rightarrow$ 0/5\\
Fresh enforce start & 0/5\\
After disable/enable (3 processes) & 5/5\\
Next switch (2 processes) & 0/5\\\addlinespace[2pt]
\tabgroup{2}{Alternatives / study variants}
Native consent (read-only): observe $\rightarrow$ switch & 1/1 $\rightarrow$ 5/5\\
Prepended registration: switch; fresh enforce & 5/5; 5/5\\
In-place update: arranged switch; fresh enforce & 5/5; 0/5\\
Volatile-mode\dg{}: arranged switch; cold observe & 5/5; 0/1\\
Single-module: observe; switch & 1/1 (3); 0/5\\
\bottomrule
\end{tabularx}
\par\medskip
\textbf{(b) OpenCode 2.2.5}\par\smallskip
\begin{tabularx}{\columnwidth}{@{}>{\raggedright\arraybackslash}Xcc@{}}
\toprule
\textbf{Allocation / condition} & \shortstack{\textbf{Selected}\\\textbf{edits}} & \shortstack{\textbf{Denied by}\\\textbf{configuration}}\\
\midrule
\tabgroup{3}{Permission Reviewer}
No saved rule: start; reload & 5/5; 5/5 & --\\
After saved allow: next prohibited edit & 0/1 (4) & --\\
After native revoke & 3/3 & --\\\addlinespace[2pt]
\tabgroup{3}{Alternatives after saved allow}
Study eligibility-only variant & 6/6 & 0/1\\
Study stage provider\dg{} & 6/6 & 0/1\\
Study check before the tool body\dg{} (1 process) & 6/6 & 1/1\\
Published ACP (1 process) & 3/3 & 1/1\\
Study ACP, renewed question (1 process) & 4/4 & 1/1\\
\bottomrule
\end{tabularx}
\par\smallskip
\parbox{\columnwidth}{\footnotesize Values are policy-evaluated/selected attempts \emph{per process}. Each condition used two processes unless parentheses give another number. Arrows show before and after an operation, and semicolons separate conditions. OpenCode workloads differ across rows, and the configuration-denied edit is additional. --: not tested. \dg{} Corrected analysis. Other routes: companion Section~V.}
\end{table}

\textbf{The same plugins and settings kept or lost participation depending on history.} Disabling and re-enabling the approval plugin through public plugin management, with plugin code and configuration unchanged, restored policy evaluation of all five selected edits, and the next switch removed it again. Processes that lost participation had the same final effective provider settings as those that kept it, but a different runtime order. No captured component check reported the loss: both providers stayed active, the policy plugin still blocked a read in enforce mode and the plugin manager raised no warning. Joining each edit to its policy record exposed the loss.\looseness=-1

\textbf{The transfer plugins lost participation the same way unless the approval question delegated.} A configuration edit or a study replay of the rewrite performed by Intutic's sync daemon recreated Intutic behind the approval plugin. In all six Intutic processes, an edit that policy prohibits executed once approved. Edit Approval on rc.8 showed the same loss after a configuration change or the switch. Through public operations alone, AgentGuard's delegating approval branch kept policy evaluating all five selected edits at the start, after its own disable and enable and after the switch, in two processes. These operations moved AgentGuard behind the policy plugin and back, so on its tested branch delegation, not order, kept policy reachable. From code, its own denial ends the chain before policy, a branch these runs did not exercise.\looseness=-1

\textbf{Moving consent after the chain kept policy in the path.} \emph{Native consent} moves approval to the host's own question after the plugin chain. Policy evaluated every selected edit through the same switch with the study controller, with live models, through public operations alone and on rc.5 (Table~\ref{tab:participation}). Four of the five API models made 54 read and edit calls, all evaluated by policy. The remaining model made no admitted call under native consent in this comparison. File edits used the non-default read-only mode, in which every edit needs the host's approval. Live models obtained wider permission for permitted edits on the first attempt or on a retry after denial. Policy participation holds only while no callback ahead of policy ends the chain (Table~\ref{tab:allocations}).

\textbf{Fixing registration order also kept policy first.} A study variant prepended the policy callback, keeping it first at an enforce start and through the switch. The \emph{readiness dependency}\dg{}, a study change, makes the approval plugin wait for the policy plugin's registered callback. In two processes it kept policy evaluating all selected edits at a start, through the switch and after reapplying the approval plugin. From code, withdrawing that readiness unloads the approval plugin, which restarts after the policy plugin re-registers.\looseness=-1

\textbf{Accepted work did not reveal whether policy took part.} The maintenance task asked live models to implement a feature in the provenance plugin's source across the switch. Acceptance required all 390 unchanged upstream tests to pass. The CLI model's work passed under both allocations. Joining host session logs to policy records gave unique decisions for all 44 post-switch covered attempts under native consent and none for the 19 under the approval plugin. One API model's work also passed under both allocations through the host's own model route, without the adapter. Policy evaluated every attempt under native consent and none of its 32 post-switch edit and shell attempts under the approval plugin. The API-adapter comparisons showed the same participation split (companion Section~VIII).\looseness=-1

\textbf{Repairs aimed at one operation failed at another.} Two study changes, an in-place update and a volatile-mode variant\dg{}, apply the switch to the running instance. Both kept an arranged policy-first order through the switch but left the approval plugin first at an unarranged cold start. Single-module repackaging of the policy plugin, a provider variant, did the reverse: it registered first in the tested starts but lost participation at the switch.

\textbf{Preventing every prohibited effect still left rejected edits unevaluated.} In one process, a study policy wrapper around dispatch, after the chain, prevented every prohibited effect in the tested sequence but evaluated only four of the six post-switch edits, missing the two that the approval plugin rejected first. It meets the GEP duty but not the every-edit requirement.

\textbf{Saved consent made edits ineligible for policy review.} On OpenCode, Permission Reviewer applies policy inside the native permission stage, before the native question, and the study configured it so that policy-allowed edits still raise that question. With no allow rule saved, it evaluated every selected edit at a start and after a reload. In four processes, one \lq\lq always\rq\rq\ reply saved an allow rule for edits to any file, which from code covers every edit tool. The next edit, which policy prohibits, took effect without evaluation, while two control processes that replied \lq\lq once\rq\rq\ evaluated every later edit. The saved rule marked the request as already allowed, so the reviewer abstained. A public issue reports this native behaviour~\cite{opencode2026issue48456}. A public revoke restored evaluation. All five API models reproduced the loss and the restoration with their own edits. The unevaluated prohibited edit breaks the GEP duty, because standing approval does not replace the policy decision. The reviewer kept its documented promise to review only asks, so the loss is a mismatch between that promise and a deployment that gives it the GEP decision.\looseness=-1

\textbf{Wider eligibility reached saved-rule edits but missed a configuration-denied edit.} The study's \emph{eligibility-only variant} of Permission Reviewer widens only its eligibility filter, to admit already-allowed edits. With the rule in place, it evaluated all six later selected edits in each of two processes. Legitimate progress and matching rejections held. Its unchanged verdict mapping still escalated policy-allowed edits, so it asked again. The \emph{stage provider}\dg{}, a study provider at that stage that leaves an allowed verdict unchanged, did not ask again. Both missed the one edit per process that OpenCode's configuration denies, which never reaches that stage.

\textbf{Checking before the tool body reached the configuration-denied edit, with different consequences for escalation.} A study check before the tool body\dg{} evaluated the configuration-denied edit. It could not ask the operator, so it turned escalations into denials. From code, the published ACP plugin also checks each call before the tool body and reuses that verdict at the permission stage, and its policy records are consistent with that reuse. Under the saved rule it kept policy deciding every selected edit, the configuration-denied one included, and prevented the prohibited edit, but its escalated edit ran without a new question. A study variant mapping ACP's escalation to the native question renewed it: a rejection prevented the edit and a one-time approval allowed it.

\textbf{A local hook remained effective after saved consent.} On Claude Code, Hookify evaluated both programmed Bash calls after an allow rule was saved and denied the matching destructive delete.\looseness=-1

\textbf{Fresh consent did not supply the designated policy decision.} In two programmed workflows, the later-selected opencode-guard asked again for every edit its rules send to the operator despite a saved allow rule~\cite{opencodeguard2026}. The guard's rejection stopped the edit. Approval allowed it unless native permission denied it. Both workflows matched all eight predictions. The local guard supplied no designated-policy decision.\looseness=-1

\textbf{Decision.} The requirement selects the placement (Table~\ref{tab:allocations}). When the chain is policy's only path, the every-edit requirement places policy before any step that can end an admitted attempt. For the unchanged pair, the next switch undid restoration, and separate fresh starts also missed policy. After either operation, operators should check that a covered edit reaches policy for evaluation.\looseness=-1

\section{Decision information}\label{sec:information}
A policy decision can be reached yet lack the source provenance that its rule needs. In DeepSeek Harness, source provenance sits in a plugin instance's private state, which reapplication discards, and in the host's conversation history, which compaction re-encodes. Table~\ref{tab:information} gives the outcomes. Controls tested restriction of edits under the study rule, or bash under the built-in rules, with a fresh untrusted source, and progress on \emph{clean work}, actions taken while no untrusted source is relevant. RiskProof's controls followed its switch, which falls by design between the source read and the restricted edit.\looseness=-1

\begin{table}[t]\centering\small
\caption{Decision information on DeepSeek Harness rc.2: restriction and clean work after each operation.}\label{tab:information}
\setlength{\tabcolsep}{2pt}\renewcommand{\arraystretch}{0.95}
\begin{tabular}{@{}>{\raggedright\arraybackslash}p{.34\columnwidth}cccc@{}}\toprule
\textbf{Allocation} & \textbf{Reapply} & \textbf{Compact} & \shortstack{\textbf{Compact,}\\\textbf{then}\\\textbf{reapply}} & \shortstack{\textbf{Clean work}\\\textbf{allowed}}\\\midrule
\tabgroup{5}{State in the plugin instance}
E Airlock & no\textsuperscript{L} & yes\textsuperscript{L}, no\textsuperscript{L} & -- & yes\\
\quad S in-place update & yes & -- & -- & yes\\
\quad S decoder fix & -- & yes, yes & no, no & yes\\
E RiskProof & no & -- & -- & yes\\
\quad S in-place update & yes & -- & -- & yes\\\addlinespace[2pt]
\tabgroup{5}{State outside the plugin instance}
S reconstruction & -- & yes, yes & yes, yes & yes\\
S journal & -- & yes, yes & yes, yes & yes\\
S registered rep.\dg{} & -- & yes, yes & yes, yes & yes\\\addlinespace[2pt]
\tabgroup{5}{No retained state}
S input-only provider & -- & yes, yes & yes, yes & no\\
\bottomrule
\end{tabular}
\par\smallskip\parbox{\columnwidth}{\footnotesize E external, S study-built. Indented rows change the plugin above. Study edit rule throughout. Operation cells report whether the restricted edit was prevented. Rows come from separate comparisons. Clean work, with no relevant untrusted source, is judged after the operations. Pairs: source summarised, then in the verbatim tail. Dash: not tested. L: also observed for a live model's own edit, same release and rule. \dg{} Analysis corrected after collection, on unchanged records.}
\end{table}

\textbf{Reapplication discards what an instance knew.} The provenance plugin Airlock labels sources by trust~\cite{dshairlock2026}. Its built-in rules restrict egress-capable tools such as bash, and a study rule that replaces them denies an edit while an untrusted source is relevant. Changing only its logging destination reapplied it, on rc.2 and alpha.2. The new instance no longer held the earlier source in its private state and treated the context as user input, so the edit that the source should restrict executed. Under the built-in rules, reapplication also released a bash call that they had blocked after a synthetic note from a Model Context Protocol (MCP) tool, although the note stayed in the model's request. This loss recurred on rc.2 and rc.5 and in a live model's own call\dg{}. RiskProof, a second provenance plugin, was set to deny untrusted edits under the host's exact-version profile exemption from its compatibility check. The study requires an observe-mode read to restrict the first enforce decision, but the switch to enforce mode reapplied the plugin and discarded that read's provenance, as its own code comment predicts~\cite{dshriskproof2026}. Each plugin owns this state, but the host's reapplication decides how long it lasts. A study in-place update, which keeps the running instance, preserved the restriction for both plugins.\looseness=-1

\textbf{A format change between releases hid a source in the verbatim tail.} Airlock promises that compaction cannot launder a label, and for a summarised source the summary inherited the label, so immediately after compaction every labelling allocation denied the edit. With the source in the verbatim tail, the original plugin permitted the edit on rc.2, on rc.5 and with a live model, although the source stayed in the model's request. The tested hosts use new field names for the replaced history range. Airlock still reads the names used by its documented host release, so its view of the history differs from the host's. From code, the plugin declares no host-release range for the host's compatibility check to enforce.\looseness=-1

\textbf{A repair that survived compaction failed at the next reapplication.} Two study repairs of Airlock kept the restriction through compaction. The \emph{decoder fix}, also tested on rc.5, reads the renamed fields but still keeps labels in the plugin instance, whereas \emph{reconstruction} rebuilds them from the host's history at each decision. A logging-only reapplication after compaction separated them: the decoder fix's new instance no longer held the source, while reconstruction kept the restriction. Four fresh repeats of these reapplication cells gave the same outcomes. The decoder fix thus addressed only the format change, while reconstruction also removed the dependence on the instance's lifetime. Reconstruction failed 41 of 390 upstream tests because the test mocks lacked the host session accessor it uses.\looseness=-1

\textbf{State outside the instance survived both operations.} A study journal persists the plugin's observed events and restores them after reapplication. Under the built-in rules on rc.5, a live model's own summary inherited the source's label. The original lost it at a reapplication, while the journal, which also carries the decoder fix, kept it. With programmed models, the \emph{registered representation}\dg{}, study-built label state that the host rebuilds from the session log, also kept the labels in nine workflows, each with up to two compactions and two reapplications. Pi's vendor plan-mode plugin, an analogous allocation, restored its restriction from the session log after a reload and a resume. By contrast, the input-only provider, which keeps no state and refuses a summary whose sources are not linked, prevented all eight restricted edits but also refused all four clean ones, because identical programmed summaries hid whether their sources were trusted.\looseness=-1

\textbf{Decision.} In the DeepSeek Harness comparisons, study-built allocations that kept state outside the plugin instance or rebuilt it from the host's history preserved restriction and clean work across the tested sequences (Table~\ref{tab:allocations}). Each repair must be tested against every operation sequence in use.\looseness=-1

\begin{table*}[!tp]\centering\footnotesize
\caption{Which tested allocations met each requirement, by operation.}\label{tab:allocations}
\setlength{\tabcolsep}{3pt}
\newcolumntype{L}[1]{>{\raggedright\arraybackslash\hangindent=0.7em\hangafter=1}p{#1}}
\newcolumntype{Y}{>{\centering\arraybackslash}p{.068\textwidth}}
\newcommand{\notmet}{\textbf{no}}
\newcommand{\rowsep}{\arrayrulecolor{black!35}\specialrule{0.4pt}{1pt}{1.5pt}\arrayrulecolor{black}}
\newcommand{\subgroup}[1]{\rowsep\firstsubgroup{#1}}
\newcommand{\firstsubgroup}[1]{\multicolumn{6}{@{}l}{\textbf{\emph{#1}}}\\}
\textbf{(a) GEP policy participation, by allocation}\par\smallskip
\begin{tabularx}{\textwidth}{@{}L{.21\textwidth}L{.19\textwidth}YYY>{\raggedright\arraybackslash\hangindent=0.7em\hangafter=1}X@{}}\toprule
\textbf{Allocation} & \textbf{Operations tested} & \textbf{Effect prevented} & \textbf{GEP duty} & \textbf{Every edit} & \textbf{Conditions and obligations}\\\midrule
\tabgroup{6}{DeepSeek Harness rc.2 unless marked}
\firstsubgroup{Approval in the same callback chain}
E Governance, E Tool Approval & Enforce start; switch after a policy-first start & \notmet & \notmet & \notmet & Reachability loss: the approval question ends the chain before policy\\
E Governance, E Edit Approval (rc.8) & Enforce start / configuration change or switch & yes / \notmet & yes / \notmet & yes / \notmet & Reachability loss: a configuration change re-registers the changed plugin last\\
\subgroup{Consent stage after policy}
E Governance, N consent after the chain & Observe start / switch & -- / yes & -- / yes & yes / yes & Read-only mode, tested file edits. No earlier callback ends the chain. Allowed edits need wider permission\\
\subgroup{Policy-first order}
S prepended registration & Enforce start; switch & yes & yes & yes & Competing prepends untested\\
S readiness dependency\dg{} & Observe start / switch, approval-plugin reapplication & -- / yes & -- / yes & yes / yes & Private field. From code, withheld readiness disables both plugins\\
\subgroup{Delegating approval}
E AgentGuard, delegating branch & Observe start, disable and enable / switch & -- / yes & -- / yes & yes / yes & From code, AgentGuard's own denial ends the chain before policy\\
\subgroup{Repairs for one operation}
S in-place update & Switch / enforce start & yes / \notmet & yes / \notmet & yes / \notmet & Held at one operation, lost at the other (reachability)\\
S volatile-mode variant\dg{} & Switch / observe start & yes / -- & yes / -- & yes / \notmet & As above\\
S single-module Governance & Observe start / switch & -- / \notmet & -- / \notmet & yes / \notmet & As above\\
\subgroup{Policy after the chain}
S wrapper around dispatch & Switch & yes & yes & \notmet & Reachability loss: two rejected edits never reached policy\\\addlinespace[0.5pt]
\tabgroup{6}{OpenCode 2.2.5}
\firstsubgroup{Policy in the permission stage}
E Permission Reviewer & Start, reload / saved allow rule & yes / \notmet & yes / \notmet & yes / \notmet & Eligibility loss: abstains on already-allowed requests, as documented\\
S eligibility-only variant; S stage provider\dg{} & Saved allow rule & yes & yes & \notmet & The configuration-denied edit never reaches the stage. The variant asks again\\
\subgroup{Policy before the tool body}
S check before the tool body\dg{} & Saved allow rule & yes & yes & yes & Escalations become denials\\
E published ACP & Saved allow rule & yes & yes & yes & An escalated edit runs without a renewed question\\
S ACP with a renewed question & Saved allow rule & yes & yes & yes & Escalations are asked again\\\bottomrule
\end{tabularx}
\par\medskip
\textbf{(b) Other hosts and relationships}\par\smallskip
\begin{tabularx}{\textwidth}{@{}L{.21\textwidth}L{.19\textwidth}L{.225\textwidth}>{\raggedright\arraybackslash\hangindent=0.7em\hangafter=1}X@{}}\toprule
\textbf{Relationship: requirement} & \textbf{Held} & \textbf{Lost (carrier question)} & \textbf{Conditions and obligations}\\\midrule
GEP local hook rule, Claude Code & Hookify: every tested Bash call after a saved allow rule & None observed & Vendor hook plugin; rule saved by a study handler. A native deny rule alone gives effect prevention only\\\rowsep
GEP decision information: reapplication and compaction & State outside the instance (Table~\ref{tab:information}). Pi plan mode, an analogue: reload, resume & Labels in the instance: reapplication (lifetime). Airlock: tail (interpretation) & Compatible source records. Registered representation\dg{}: synchronous, pure updates (from code), which the study requires not to throw\\\rowsep
AC response applied: before progression & Per-call review: all nine workflows, normal and direct routes & Batched review: direct route (reachability) & Documented limits on middleware order, resumption and answers covering several calls\\\rowsep
AC completed work: none repeated & Per-call tasks: resumption & Study grouped node: resumption (lifetime) & A saved boundary that resumption reuses\\\rowsep
Blackboard membership: exclusion until re-enrolment & Study repair of lookup and status: restart and running lookup, on two releases & Ensemble: running lookup (lifetime) & Two changes tested together. If resumption re-enrols, Ensemble's delivery is correct\\\rowsep
Restorations (GEP, Blackboard) & Disable and enable, revoke (OpenCode), remove and restore (Ensemble) & The next switch undid disable and enable (reachability) & Tested between calls. Re-check after relevant operations\\\bottomrule
\end{tabularx}
\par\smallskip\parbox{\textwidth}{\footnotesize N native, E external, S study-built. In enforce mode the requirements nest (Section~\ref{sec:background}). Yes: the assessed actions met the requirement under the listed operations and conditions. The GEP-duty column covers policy participation and enforcement of prohibitions, and conditions report escalation handling. Slashes follow the listed operation order. At observe starts only evaluation is assessed, so effect and GEP-duty cells are dashed. Carrier questions classify observed losses. Counts: Table~\ref{tab:participation}, Sections~\ref{sec:participation}--\ref{sec:approval-membership}. \dg{} Analysis corrected after collection on unchanged records (Section~\ref{sec:design}).}
\end{table*}

\section{Approval, resumption and membership}\label{sec:approval-membership}\label{sec:continuation}
Three supporting requirements, each under its own authority (Table~\ref{tab:requirements}), test execution routes and retained state beyond GEP. On each tested route the operator's response must apply before its pending action progresses. The maintenance task requires resumption to avoid repeating completed work. The Blackboard deployment requires a shut-down member to stay excluded until it re-enrols, both while the board plugin runs and after a restart.\looseness=-1

\textbf{Per-call review held on every tested route, and batched review failed on direct routes.} Eighteen programmed Deep Agents workflows crossed two review modes, three routes and three operator responses (approval, rejection and none). LangChain's batched review asks about a model turn's calls together and records a rejection only as a message. The default router then withholds the rejected call on the normal route to the \emph{tool node}, the graph step that runs tools. Per-call review applies the response where the call executes, before its effect. Both review modes held on the normal route. A direct route to the tool node, which LangChain middleware can request, skips the router, as a user-reported issue records~\cite{langchain2026reviewroute}, and a study middleware took it before or after review. Per-call review held in all nine of its workflows, six of them on the direct route. Batched review violated the requirement in four of the six direct-route workflows: three calls executed before any question when the middleware jumped before review, and one executed after its delivered rejection. Deep Agents' approval option builds batched review and reproduced the execution after a delivered rejection in both its processes. This is a \emph{composition loss}: a decision protects only the routes that consult it.\looseness=-1

\textbf{Pending requests survived where their continuation state did.} On Deep Agents, graph recreation with the same checkpoint store and thread kept the pending call, and its matching response then applied. An empty store lost the pending work, and an approval for the original request, sent on another thread, left both threads' requests pending. On OpenCode the server kept a pending request across client reattachment.\looseness=-1

\textbf{Resumption avoided repeating completed work where each result was saved.} In a programmed Deep Agents maintenance task built from Airlock's source and tests, each pending edit followed its matching approval or rejection. Ordinary \emph{per-call tasks}, which run each tool call separately, save each completed result, so resumption did not repeat a completed configuration edit, and after approval the project passed all 390 upstream tests. A study-built node grouped the completed edit with the pending one, an arrangement LangGraph's documentation warns against~\cite{langgraph2026interrupts}. Resumption repeated the completed edit after either reply. After approval the duplicate change prevented compilation.\looseness=-1

\textbf{The same stored status preceded opposite membership decisions.} The board plugin documents that its lookup returns only members that have not shut down~\cite{ensemble2026}. From code, its running lookup reads a member cache with no expiry, whereas after re-initialisation a check of stored status decides. Membership thus has two sources of truth, although no runtime trace shows which lookup ran. In each of two matched pairs, the prerequisite board calls and arguments matched by role, and both runs stored the member as shut down. Only one run per pair then re-initialised the plugin: by restart, or by removing and restoring it through native configuration in the same host process. Without re-initialisation, the shut-down member's read delivered the new item into its next model request. After re-initialisation, the read was refused and the item appeared in neither the tool result nor that request. The continuing member received the item in both arms. Under the study requirement that re-enrolment must be explicit, the running lookup violates both the provider promise and the deployment requirement. If a resumed prompt instead re-enrols the member, delivery is correct and refusal after re-initialisation blocks the expected resumption. A study repair kept the exclusion in the running process and across a restart on two plugin releases. It makes the running lookup also check stored status and keeps a shut-down member shut down when its status updates. We tested both changes together.\looseness=-1

\section{Synthesis across cases}\label{sec:discussion}

\textbf{RQ1: allocations and carriers.} Responsibility for a relationship and control of its carrier could lie in different providers. The host controlled registration, reapplication, permissions and history format in the principal-case losses, while the board plugin controlled its own member cache. Connector order or position carried policy participation and AC review. Retained state carried decision information, completed work and membership, and saved consent decided reviewer eligibility. The saved-rule loss occurred while Permission Reviewer followed its documented eligibility rule, so a provider can meet its promise while the allocation fails a stronger requirement.\looseness=-1

\textbf{RQ2: operations and mechanisms.} Preservation depended on operational history and sequence. With the same plugins and final settings, different histories gave different policy coverage, and a restoration lasted only until the next switch. The decoder fix kept provenance through compaction but lost it at the next reapplication, so success under one operation does not establish preservation through a sequence. Supporting cases exposed carrier dependencies: a direct route skipped the router that applies the AC response, a grouped node repeated completed work and re-initialisation restored exclusion under the study's explicit re-enrolment requirement.\looseness=-1

\textbf{RQ3: allocations, conditions and trade-offs.} The requirement determined which allocations were acceptable (Table~\ref{tab:allocations}). Three reachability tactics preserved participation on DeepSeek Harness under their tested conditions: fixed policy-first order, consent after the chain and delegating approval. Effect prevention and the GEP duty admitted allocations that missed edits rejected elsewhere, such as the wrapper around dispatch and wider eligibility on OpenCode, whereas the every-edit requirement excluded them. Evaluation alone did not determine escalation handling. The check before the tool body\dg{} denied escalations, while published ACP executed an escalated edit without a new question. The study ACP variant requested fresh consent. For decision information, state kept outside the plugin instance or rebuilt from the host's history held through the tested sequences.\looseness=-1

Preserving allocations introduced different costs: wider permission requests under native consent, renewed questions from the eligibility-only variant, a private field for the readiness dependency and upstream test failures for reconstruction.\looseness=-1

\textbf{Five carrier questions.} From the cases we derive five overlapping questions for assessing an allocation: four concern carrier failures and one concerns control. \emph{Reachability}: is the required decision reached before a covered action executes and, where every attempt must be evaluated, also for attempts that another provider ends? Its first part restates complete mediation for a chain of providers. \emph{Eligibility}: does the provider act on every request state that the requirement covers? \emph{Lifetime}: does the state a decision reads stay available and current for as long as the decision needs it? It covers state transfer when an update replaces state needed by a later decision. \emph{Interpretation}: do readers and writers agree on format and meaning? Disagreement is a form of architectural mismatch. \emph{Control}: who sets or changes the carrier's order, lifetime, consent state or representation, and is it the provider that relies on it?\looseness=-1

\textbf{Implications.} For architects, choosing an allocation means asking whether the operations in use can rewrite its carrier. In the DeepSeek Harness chain a terminal ask or allow skips any policy behind it, much like first-applicable combination~\cite{xacml2013} but under an undeclared order. Harness designers can declare precedence or dependencies instead. Deny-overrides, as among Kubernetes validating admission webhooks~\cite{k8sadmission}, prevents an effect when a reached policy prohibits it. If a callback chain is policy's only path, evaluating every admitted attempt requires policy before any step that can end the chain. Pattern catalogues can name each relationship's carrier and the operations that may rewrite it. Assessors need records joined per action, since task success, plugin health and effect checks alone do not show whether policy evaluated an attempt.\looseness=-1

\section{Related work}\label{sec:related-work}
\textbf{Patterns and conformance.} Conformance checking compares intended and implemented architecture~\cite{murphy2001,passos2010,yan2004discotect}, also through model-based tests and run-time checks~\cite{uzun2019conformance,ganesan2009runtime}. Pattern contracts and event-based monitors check patterns at runtime and during reconfiguration~\cite{hallstrom2004monitoring,krieger2020}. AgentArcEval evaluates agent architectures using scenarios informed by stakeholder input~\cite{lu2025agentarceval}. Our comparisons exercise counterparts of two harness patterns that Barbaste et~al.\ identify~\cite{barbaste2026}: Middleware Pipeline (LangChain middleware and the tool callback chain) and Event Sourcing (state derived from the session log).\looseness=-1

\textbf{Composition and agent controls.} Zhang et~al.\ find unsafe combinations of individually safe updates to harness components~\cite{zhang2026csf}. Cordis manages declared dependencies reactively~\cite{shi2026cordis}. Our callback-order losses arose between two baseline plugins that declare no dependency on each other. Quiescence~\cite{kramer1990} and state transformation~\cite{hayden2012kitsune} support dynamic reconfiguration. In our cases, reapplication discarded provenance held in the replaced instance without such transfer. Loopjacking tests how approval binds to the executed action~\cite{kumar2026loopjacking}, whereas our participation comparisons test whether operations keep policy reachable. Zheng and Yang specify assume--guarantee contracts that carry provenance and action binding across composed agent controls~\cite{zheng2026continuity}, and AGATE decides provenance-based authorisation in one deterministic core across DeepSeek Harness, OpenCode and OpenClaw~\cite{zhang2026agate}. Khan's machine-checked resume contract forbids repeating an effect on the same branch, as tested by our completed-edit contrast~\cite{khan2026resume}. Our contribution is empirical: among existing native services and independently written plugins, the same providers and final settings gave different policy coverage after different operation histories, and whether a repair held depended on the operation sequence.\looseness=-1

\section{Threats to validity}\label{sec:threats}
\textbf{Construct.} Requirements, code readings and expectations were derived from the pattern sources, provider documentation and pinned code with AI assistance, and separate analyses checked consequential readings, bindings and judgments against source texts and native records~\cite{baltes2025guidelines}. Some requirements are study choices, such as the every-edit requirement, RiskProof's carry-over and Blackboard's explicit re-enrolment, and another choice changes which judgments hold, not the observed effects. The prohibited edits that executed after the switch and the saved-rule operation establish loss of the GEP duty as well as the every-edit requirement. The switch, native-consent and decoder-fix versus reconstruction comparisons provide evidence that depends neither on the RiskProof and registered-representation\dg{} cases nor on any corrected analysis.\looseness=-1

\textbf{Internal.} Operator responses were programmed throughout, as were model responses outside the live-model comparisons, which repeated four GEP losses and the native-consent preservation in the tested tasks. The original validation of joins from recorded fields alone used one shared clock and trajectories without repeated calls sharing session, tool and complete arguments. Starts through public operations alone, with no study code in the host, showed the startup loss in all 32 runs, two earlier starts and ten each with no added load, CPU load or direct I/O. Every operation in Table~\ref{tab:participation} except the \lq\lq always\rq\rq\ reply ran with no edit in flight, so the judgments concern operations between calls.\looseness=-1

\textbf{External.} The switch and composition with other providers are documented as intended practice~\cite{dshgovernance2026,agentguard2026}. The purposively selected cases show that these losses and preserving allocations occur in pinned released assemblies, not how often (companion Section~IX). Participation evidence rests on one DeepSeek Harness pair, extended by the transfer cases, two OpenCode policy providers and one Claude Code hook plugin. Chain-ending approval plugins cause the loss only for the tools they cover and when they run before policy. The studied DeepSeek Harness releases are developer previews, and on rc.5 the published plugin archives were staged directly. Later Deep Agents and LangChain releases were compared from code only.\looseness=-1

\section{Conclusion}\label{sec:conclusion}
The study shows that required relationships can fail while all participating providers remain active. In the principal case, the same plugins and settings gave different policy coverage after different operational histories, and a repair that survived one operation failed at the next. Allocations that kept the required order, or kept or rebuilt the needed state, held on their tested paths and operation sequences, and preventing prohibited effects admitted more allocations than evaluating every admitted edit. Architects should identify each relationship's carrier, who controls it and which operations can change it, then test the chosen allocation against the routes and operation sequences in use.\looseness=-1

\textbf{Acknowledgements.} Codex and Claude aided writing, coding and analysis. The authors checked and remain responsible.

\bibliographystyle{IEEEtran}
\bibliography{refs}

\end{document}